# Omnidirectional MediA Format (OMAF): Toolbox for Virtual Reality Services

Sachin Deshpande
Sharp Labs of America
Camas, WA, USA
sdeshpande@sharplabs.com

Miska M. Hannuksela
Nokia Technologies
Tampere, Finland
miska.hannuksela@nokia.com

***Abstract*—This paper provides an overview of the Omnidirectional Media Format (OMAF) standard, second edition, which has been recently finalized. OMAF specifies the media format for coding, storage, delivery, and rendering of omnidirectional media, including video, audio, images, and timed text. Additionally, OMAF supports multiple viewpoints corresponding to omnidirectional cameras and overlay images or video rendered over the omnidirectional background image or video. Many examples of usage scenarios for multiple viewpoints and overlays are described in the paper. OMAF provides a toolbox of features, which can be selectively used in virtual reality services. Consequently, the paper presents the interoperability points specified in the OMAF standard, which enable signaling which OMAF features are in use or required to be supported in implementations. Finally, the paper summarizes which OMAF interoperability points have been taken into use in virtual reality service specifications by the 3rd Generation Partnership Project (3GPP) and the Virtual Reality Industry Forum (VRIF).**

## I. INTRODUCTION

The world around us is three-dimensional (3D) and immersive. Yet, a majority of traditional media we consume has been two-dimensional (2D). 360° video, audio and associated media can be consumed on a stereoscopic head-mounted display to obtain an immersive media experience which is often far superior to the 2D media consumption experience. To allow interoperability for such a virtual reality (VR) ecosystem, the Moving Picture Experts Group (MPEG) has developed a new standard: ISO/IEC 23090-2 Omnidirectional Media Format (OMAF). The 1st edition of the standard [1] was completed in October 2017 and specified the format for 360° video, images, and audio intended for three degrees of freedom (3DOF) rendering. The 2nd edition of the standard [2] was finalized in October 2020 and added support for multiple viewpoints and overlays. A viewpoint represents omnidirectional media corresponding to one omnidirectional camera, and an overlay is defined as media that is superimposed over omnidirectional background video or image or on the viewport. The 2nd edition also supports six degrees of freedom (6DOF) rendering in such a way that the movement of the viewing position impacts the rendering of overlays in relation to the omnidirectional background video or image.

This paper gives an overview of OMAF with an emphasis on multiple viewpoints, overlays, and features that can be used as interoperability points in VR services. This manuscript discusses the usage scenarios for multiple viewpoints and overlays more extensively than in any of the earlier papers [3][4]. We hope that the discussion enlightens professionals in various fields to understand the potential of OMAF for interchange of immersive VR content. OMAF is not a monolithic standard but offers a toolbox of features. Consequently, the paper provides an understanding of the interoperability points provided by OMAF and how they can be applied in different domains. We review VR service standards and specifications that reference the interoperability points specified by OMAF. As far as we know, such a review has not been provided earlier.

The remainder of this paper is organized as follows. Section II provides an overview of OMAF, including the architecture, the referenceable features in the OMAF toolbox, and a high-level overview of potential usage scenarios for OMAF. Sections III and IV describe support for multiple viewpoints and overlays, respectively. Section V reviews the VR service standards and specifications that reference and use OMAF. Finally, Section VI concludes the paper.

## II. OVERVIEW OF OMAF

### *A. Architecture*

Fig. 1 shows a typical content flow for an omnidirectional media application. The pre-processed pictures (D) are encoded as coded images ($E_i$) or a coded video bitstream ($E_v$). The captured audio ($B_a$) is encoded as an audio bitstream ($E_a$). The coded images, video, and/or audio are then composed into a media file for file playback (F) or a sequence of an initialization segment and media segments for streaming ($F_s$). The segments $F_s$ are delivered using a delivery mechanism to a player. The file encapsulator output file (F) and file decapsulator input file (F') are identical. A file decapsulator processes the file (F') or the received segments ($F'_s$) and extracts the coded bitstreams ($E'_a$, $E'_v$, and/or $E'_i$) and parses the metadata.

Viewport-dependent video may be carried in multiple tracks, which may be merged in the bitstream rewriting (a.k.a. tile binding) into a single video bitstream $E'_v$ prior to decoding. The audio, video, and/or images are then decoded into decoded signals ($B'_a$ for audio, and D' for images/video). In the image rendering block, the decoded pictures (D') are projected onto the screen of a head-mounted display or any other display device based on the metadata parsed from the file. Similarly, decoded audio ($B'_a$) is rendered, e.g. through headphones, according to the current viewing orientation. Viewing orientation tracking function which may use head and/or eye tracking determines the current viewing orientation. When sphere-relative overlays are in use, the viewing orientation tracking functionality could include or be complemented by viewing position tracking and rendering of overlays with background visual media can take both the viewing position and the viewing orientation into account. In viewport-dependent delivery, the current viewing orientation is passed to the strategy module, which determines the video tracks to be received based on the viewing orientation.

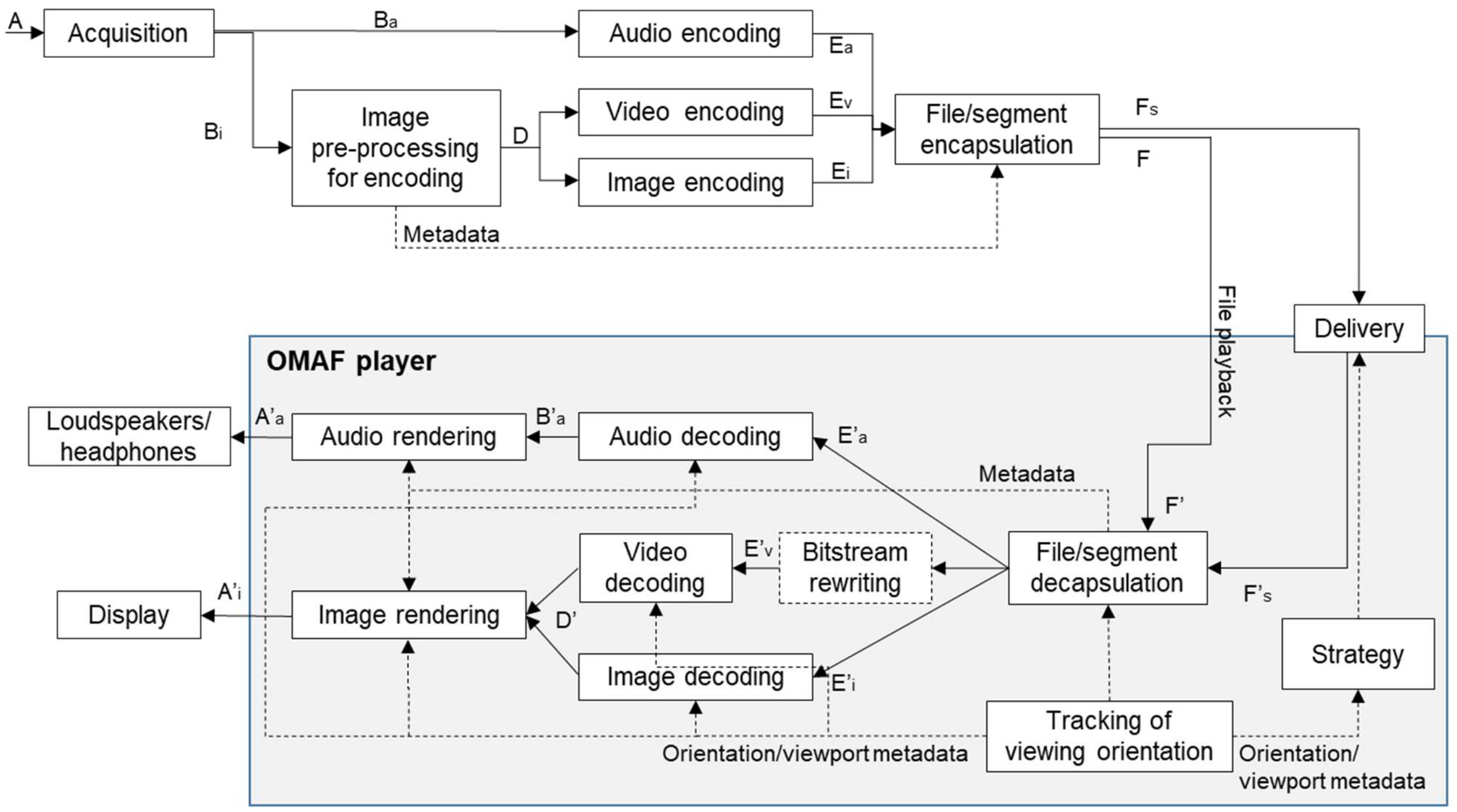

Fig. 1. Main components of omnidirectional media application [2]

Viewport-dependent streaming reduces the bitrate and can be used to drop the decoding capacity requirement without sacrificing the resolution on the viewport. Viewport-dependent streaming can be achieved with two types of methods: viewport-specific streams or tile-based streaming. Viewport-specific streams area tailor-made to specific viewing orientations, where the quality or resolution emphasis on the viewport could be achieved by various means. In tile-based viewport-dependent streaming, the video is split into tiles, which may be encoded at multiple resolutions and/or qualities. Tile binding may be carried out in the player to combine the selectively received tiles into a single bitstream for decoding. Please refer to [3] for a more extensive overview of viewport-dependent streaming.

OMAF focuses on specifying following interfaces:

- $E'_a$, $E'_v$, $E'_i$: audio bitstream, video bitstream, coded image(s), respectively.
- F/F': media file and media profiles specified in include the specification of the track formats for F/F'. The file format features for timed media of OMAF build on the ISO base media file format (ISOBMFF) [5], whereas OMAF extensions for images conform to the High Efficiency Image File Format (HEIF) [6].
- $F_s$/$F'_s$: The delivery related interfaces for delivery using the Dynamic Adaptive Streaming over HTTP (DASH) [7] or the MPEG media transport (MMT) [8].

### B. *Usage Scenarios for OMAF*

The OMAF toolbox of features facilitates a broad variety of usage scenarios, some of which are reviewed in this subsection. Specific usage scenarios for multiple viewpoints and overlays are discussed in Section III and IV, respectively.

Smartphones or cameras with fisheye lens(es) could store captured images using HEIF [6]. The HEIF files can contain OMAF image metadata so that they can be reproduced correctly in any rendering device. Images can be stored either in the fisheye format or stitched to a equirectangular projection (ERP) or cubemap projection (CMP) image. HEIF facilitates the use of efficient codecs, such as HEVC, providing higher fidelity (10 bits per color component and high dynamic range) and significantly better compression than JPEG [9].

OMAF provides standardized metadata that enables omnidirectional audio-video file capturing and playback by extending the broadly used ISOBMFF. Both monoscopic and stereoscopic video are supported, and any content coverage up to 360° can be signaled.

OMAF specifies segment formats and DASH Media Presentation Description (MPD) metadata for 360° video streaming. OMAF supports both viewport-independent and viewport-dependent streaming. Some deployments can choose to use viewport-independent streaming, which is simpler but less efficient in terms of rate-distortion performance compared to the viewport-dependent streaming.

OMAF enables interactive multimedia presentations through multiple viewpoints and overlays. OMAF can be used as the standardized interchange format for authoring and distributing such presentations. It is remarked that the interaction features of OMAF can be used for conventional two-dimensional video too, e.g. for realizing multiple user-selectable storyline paths.

While OMAF was primarily designed for virtual reality, it also provides basic support for augmented reality (AR), where viewpoints can be aligned with geographical position and geomagnetic coordinates.

### C. *Feature Toolbox*

OMAF provides a toolbox of features, which can be selectively applied in applications, services, and external specifications. OMAF aims at facilitating interoperability of omnidirectional media across content authoring tools and player applications. Thus, OMAF specifies referenceable features at different levels of granularity. This subsection reviews the concepts and referenceable code points that applications and external specifications can selectively use from the OMAF standard.

OMAF specifies the following **types of omnidirectional video and images**:

- Projected omnidirectional video/images: the relation of a spherical video/image to a respective two-dimensional video/image is described through a mathematically specified projection.

- Fisheye omnidirectional video/images: one or more circular images originating from fisheye lenses and captured simultaneously are spatially arranged onto a picture.

- Mesh omnidirectional video: rectangular regions of two-dimensional pictures are mapped to mesh elements of a three-dimensional mesh, where mesh elements can be either parallelograms or sphere regions corresponding to rectangular areas in equirectangular projection.

**Media profiles** specify how a media codec is adapted for omnidirectional application usage. An OMAF media profile is defined as requirements and constraints on media coding as well as on signaling and encapsulation of the media data in an ISOBMFF file. OMAF specifies media profiles for video, images, audio, and timed text, as summarized in TABLE I, II, III, and IV, respectively.

The OMAF video profiles use either the High Efficiency Video Coding (HEVC) standard [10] or the Advanced Video Coding (AVC) standard [11] as basis. There are two HEVC-based viewport-independent profiles, one for resolutions up to 4K (e.g., 4096×2048) and another for unconstrained resolutions. The latter profile was added in OMAF 2nd edition to align with the increased decoding capacity and display resolutions of VR playback devices for which 8K (e.g., 8192×4096) is justified. The HEVC-based and AVC-based viewport-dependent OMAF video profiles support both viewport-specific streams and tile-based streaming. These profiles require the content author to generate a prescription of the tile binding operation through an extractor track. The simple and advanced tiling OMAF video profiles were added into OMAF 2nd edition. They provide a greater flexibility for player-driven tile selection as opposed to the prescribed tile selection that only allows players to select between different quality versions of tiles per a single extractor track. Please refer to [3] for a more extensive overview of the tile binding approaches and OMAF video profiles.

**Toolset brands** indicate functionalities beyond basic playback of omnidirectional audiovisual content. OMAF 2nd edition includes toolset brands for multiple viewpoints, non-linear storyline, and overlays.

OMAF specifies **scheme types for restricted video tracks**. The scheme types i) indicate one of projected, fisheye, or mesh omnidirectional video being used; and ii) may constrain which rendering features are needed.

OMAF specifies metadata for characterizing omnidirectional video or image. Moreover, to realize certain additional functionalities for playback and viewport-dependent content selection, OMAF also specifies timed

TABLE I. OMAF MEDIA PROFILES FOR VIDEO

| Media Profile | Codec | Level | Type |
|---|---|---|---|
| HEVC-based viewport-independent OMAF video profile | HEVC Main 10 | 5.1 | Projected with ERP |
| Unconstrained HEVC-based viewport-independent OMAF video profile | HEVC Main 10 | any | Projected with ERP |
| HEVC-based viewport-dependent OMAF video profile | HEVC Main 10 | 5.1 | Projected with ERP or CMP |
| AVC-based viewport-dependent OMAF video profile | AVC Progressive High | 5.1 | Projected with ERP or CMP |
| Simple tiling OMAF video profile | HEVC Main 10 | any | Projected with ERP or CMP |
| Advanced tiling OMAF video profile | HEVC Main 10 | any | Mesh |

TABLE II. OMAF MEDIA PROFILES FOR IMAGES

| Media Profile | Codec | Level |
|---|---|---|
| OMAF HEVC image profile | HEVC Main 10 | 5.1 |
| OMAF legacy image profile | JPEG | Not applicable |

TABLE III. OMAF MEDIA PROFILES FOR AUDIO

| Media Profile | Codec | Profile | Level | Max Sampling Rate |
|---|---|---|---|---|
| OMAF 3D audio baseline profile | MPEG-H Audio | Low Complexity | 1, 2 or 3 | 48 kHz |
| OMAF 2D audio legacy profile | AAC | HE-AACv2 | 4 | 48 kHz |

TABLE IV. OMAF MEDIA PROFILES FOR TIMED TEXT

| Media Profile | Codec | Profile |
|---|---|---|
| OMAF IMSC1 timed text profile | IMSC1 | Text Profile or Image Profile |
| OMAF WebVTT timed text profile | WebVTT | n\a |

metadata tracks, boxes, and track groups, which are described as follows:

Region-wise quality ranking (RWQR) information provides relative values of the quality of indicated sphere regions or 2D regions on a decoded picture. The information can be used for content selection for viewport-dependent streaming.

ERP region timed metadata provides time-varying relative quality rank recommendation, relative priority information, or heatmap signaling for a rectangular grid relative to ERP. Players can use this metadata for region-adaptive bitrate adaption choices in tile-based streaming.

Initial viewing orientation timed metadata provides the viewport direction for random accessing and resetting the viewing orientation towards to content author's choice after a scene cut. The respective property for images gives the initial viewing orientation to be used for image rendering.

Recommended viewport timed metadata provides the time-varying viewport chosen by the content author or created through viewing statistics. This metadata enables a playback mode where rendering follows recommended viewing orientations and positions.

2D spatial relationship track grouping describes the spatial layout of sub-picture tracks. It can be used for viewport-dependent streaming, particularly when multiple video decoder instances can be executed in parallel.

## III. Multiple Viewpoints

### A. Usage Scenarios for Multiple Viewpoints

Content with multiple viewpoints can be authored in a manner that enables users to navigate freely between viewpoints or restricted so that users follow pre-defined storyline paths with controlled freedom of selecting between upcoming viewpoints.

Multi-viewpoint content with free user navigation can be used e.g. for virtual museums, natural parks, historic sights, city walks, sports, talk shows, and concerts. As an example, a detailed description of Intel's solution for multi-camera virtual reality sports production is available in [12], section D.2. OMAF overlays have been demonstrated together with multiple viewpoints in a city walk use case in [13]. In many productions directors may prescribe the camera switches, but users are given the option to select cameras themselves.

OMAF allows the viewpoints to have Global Positioning System (GPS) coordinates and have the alignment of the viewpoint coordinate system with the geomagnetic north to be signaled. Consequently, it is possible to use OMAF for outdoor augmented reality. For example, the viewpoint can be selected based on the actual GPS position of the device. This can be used for providing additional video, audio, image, and description for outdoor historic sites or city walks at the actual geographic location and orientation.

Multi-viewpoint content with selectable storyline paths can be realized by defining each storyline segment as a viewpoint and thus used e.g. for the following usage scenarios:

- Training material: OMAF can be used to create linear or non-linear paths through the material. The material can contain omnidirectional and/or two-dimensional video and images as well as audio. Text, image and video overlays can be used to enrich the content and to provide user interaction, such as questionnaires.
- Interactive omnidirectional or two-dimensional films: The user is every now and then presented with options on how to proceed. The user might have a limited time to pick an option. After the limited time has elapsed, a default choice of the next viewpoint may be automatically selected.

### B. OMAF Signaling for Multiple Viewpoints

The viewpoint ISOBMFF information structures provide the following metadata information:

- Viewpoint position information: viewpoint position (x, y, z) relative to other viewpoints and GPS position.
- Viewpoint orientation: rotation angles of the global coordinate system of the viewpoint relative to the common reference coordinate system across all viewpoints, and the orientation of the common reference coordinate system relative to the geomagnetic North direction.
- Viewpoint grouping information: allowing specification of logical grouping of viewpoints.
- Viewpoint switching information: including how timelines are handled when switching between viewpoints, regions that initiate viewpoint switching under user control.
- Viewpoint looping information: allowing playback repetition control. This can be used when waiting for user's selection on which viewpoint is played next.

The information about multiple viewpoints signaled in OMAF may be either statics or dynamic:

- Static information: For many typical use cases the viewpoint (i.e. omnidirectional cameras) may be stationary throughout the video sequence, in this case it is sufficient to signal viewpoint information statically – one time.
- Dynamic information: For certain use cases, such as sports, concerts, or other events, one or more viewpoints may be moving. In this case OMAF metadata supports time varying viewpoint information signaling using dynamic viewpoint timed metadata tracks.

The OMAF viewpoint information (VWPT) DASH descriptor in a Media Presentation Description (MPD) allows a client to select the appropriate viewpoint and switch between multiple viewpoints seamlessly with low latency. The VWPT descriptor directly provides information that is similar to the viewpoint ISOBMFF metadata.

## IV. Overlays

### A. Usage Scenarios for Overlays

Overlays can be used in, but are not limited to, the usage scenarios presented in TABLE V.

TABLE V. Examples Usage Scenarios for Overlays

| **Category** | **Description and Examples** |
|---|---|
| Annotation | An overlay that annotates the content. Examples:<br>1. A score box or player statistics superimposed on a live omnidirectional game (which is background media).<br>2. A textual ticker, e.g. for news.<br>3. Semi-transparent annotation maps, where each user-switchable overlay could highlight and annotate a different aspect of the background image or video. |
| Logo | A still image or a repetitive animation. Logo overlays can take advantage of tuning the opacity level with alpha planes. E.g. content provider's logo. |
| Watermark | A semi-transparent watermark on top of the content. E.g. the name of the copyright holder. |
| Advertisement | Viewport-relative advertisements are displayed regardless of the viewing orientation. Sphere-relative advertisements can relate to the background content. |
| Closeup | Two-dimensional video or image closeups of the omnidirectional video or image on the background. Such closeups may offer replays of an event in a sports game, for example. A specific overlay source is the recommended viewport for the content. |
| User control | Overlays that enable users to interact with the content. Examples:<br>1. Hotspots for switching viewpoints.<br>2. Buttons or icons for turning video/image/text overlays on or off. |
| Application content | Content from any external source or application can be included as an overlay in an OMAF presentation. E.g., a conventional two-dimensional video stream can be embedded into a viewing space with 360° background image or video and possibly with other overlays. |

## B. *OMAF Signaling for Overlays*

In OMAF an overlay is specified with the following pieces of information:

- overlay source, specifying which track or image item and, in some cases, which spatial region within the track or the image item are used as the content of the overlay
- rendering type of the overlay, specifying whether the overlay is anchored relative to the viewport or to the unit sphere and, for sphere-relative overlays, the shape of the surface on which the overlay is rendered (plane or sphere region)
- rendering properties of the overlay, specifying e.g. the opacity of the overlay
- user interaction properties for the overlay

The overlay source is one of the following: i) entire decoded pictures of a video track; ii) an entire output image of an image item; iii) an indicated rectangular overlay source region within decoded pictures of a video track or an output image of an image item; iv) a recommended viewport of omnidirectional video, as indicated by a recommended viewport timed metadata track; or v) an externally specified overlay source. When a source region is used as overlay source, the picture that contains the source region may also contain other overlay source regions or background visual media.

The rendering type of the overlay is one of the following: i) viewport-relative overlay, specifying that the overlay is displayed on a rectangular area at a specified position relative to the viewport; ii) sphere-relative projected omnidirectional overlay, specifying that the overlay is displayed on a sphere surface at an indicated position within or on the unit sphere; iii) sphere-relative 2D overlay, specifying that the overlay is displayed on a plane at an indicated position within the unit sphere; or iv) 3D mesh overlay, specifying that the overlay is displayed on indicated mesh elements of the 3D mesh.

The following rendering properties may be indicated for an overlay: i) layering order, specifying the layering order among the overlays that are relative to the viewport, and separately among each set of overlays that have the same distance from the center of the unit sphere; ii) opacity of the overlay that is applied for all pixels of the overlay source when rendering the overlay; iii) priority value, based on which a priority order of overlays can be derived and used in the case that an OMAF player does not have enough decoding capacity to decode all overlays; value 0 indicates that the overlay is essential for displaying; and iv) alpha composition, specifying that the overlay is associated with an alpha plane used for determining pixel-wise opacity and blending when superimposing the overlay.

The following user interaction properties may be indicated for an overlay: i) controls which user interactions are allowed in a user interface; ii) overlay label, which could be used in a user interface; and iii) associated sphere region, intended to be used as a user-selectable area to turn an overlay on or off.

Overlay controls can be static or time-varying. Static overlay controls are contained in a sample entry in a track containing the overlay source or as an item property for an image item containing the overlay source. Time-varying overlay controls are included in a timed metadata track.

An overlay can be either active or inactive. An inactive overlay shall not be displayed. In addition to the active/inactive status for overlays, an overlay has a state on or off, which indicates whether an overlay has been turned on or off by a user interaction. When an overlay is turned off by a user interaction, it is not displayed.

The OMAF overlay information (OVLY) DASH descriptor specifies overlays in DASH Representations associated with the descriptor. The OVLY descriptor allows distinguishing, directly from the MPD file, between Adaptation Sets that contain overlays from the Adaptation Sets that contain background visual media. The OVLY descriptor provides following information about overlays in an MPD file: i) a mandatory list of overlay ID values for overlays; and ii) an optional list of priority values for the overlays listed in the above list.

Additionally OMAF allows defining viewing space intended to be used for head-tracked rendering of background visual media and overlays.

# V. Use of OMAF in Other Specifications

## A. *3GPP Specification on VR Profiles for Streaming Applications*

The 3rd Generation Partnership Project (3GPP) standardizes cellular telecommunications, including multimedia services. 3GPP Technical Specification (TS) 26.118 on VR profiles for streaming applications [14] defines operation points and media profiles for video and audio. The operation points specify constraints on the coding format and the audio or video signal. A media profile specifies which operation point is used and how a bitstream conforming to the operation point is encapsulated in a file and streamed over DASH. VR media profiles by 3GPP are intended to serve as interoperability points in VR streaming services and devices.

The VR video operation points by 3GPP specify spatial and temporal resolutions, color spaces, transfer functions, projection formats, video codecs, and the profile, tier, and maximum level of the video codecs. 3GPP TS 26.118 specifies four VR video operation points, namely Basic H.264/AVC, Main H.265/HEVC, Main 8K H.265/HEVC, and Flexible H.265/HEVC, which are summarized in TABLE VI. The Basic H.264/AVC and Main H.265/HEVC operation points only allow standard dynamic range, whereas the Main 8K and Flexible H.265/HEVC operation points also support high dynamic range. The Basic H.264/AVC operation point only allows ERP with full 360° coverage, while the Main and Main 8K H.265/HEVC operation points support ERP with any content coverage, and the Flexible H.265/HEVC operation point supports both ERP and CMP with any content coverage and unlimited region-wise packing.

3GPP TS 26.118 specifies three VR video profiles, namely Basic Video, Main Video, and Advanced Video, which respectively use the Basic H.264/AVC, Main or Main 8K H.265/HEVC, and Flexible H.265/HEVC operation point. The file encapsulation of 3GPP's VR video profiles is based on the OMAF-specified scheme types. The Basic Video profile only supports viewport-independent streaming. The Main Video profile also supports viewport-specific streams, which can be signaled through the OMAF-defined sphere region quality ranking descriptor or 3GPP-specific MPD signaling. The Flexible Video profile adds also tile-based viewport-dependent streaming support similarly to HEVC-based viewport-dependent OMAF video profile.

TABLE VI. VR Video Operation Points by 3GPP

| Operation point | Codec | Level | Projection format | Stereo |
|---|---|---|---|---|
| Basic H.264/AVC | AVC High | 5.1 | ERP w/o padding | No |
| Main H.265/HEVC | HEVC Main 10 | 5.1 | ERP w/o padding | Yes |
| Main 8K H.265/HEVC | HEVC Main 10 | 6.1 | ERP w/o padding | Yes |
| Flexible H.265/HEVC | HEVC Main 10 | 5.1 | ERP w/o padding or CMP | Yes |

3GPP TS 26.118 specifies a single VR audio operation point and a single audio profile, which are aligned with the definition of the OMAF 3D audio baseline profile.

### B. *VR Industry Forum Guidelines*

The VR Industry Forum (VRIF) has the mission to advocate industry consensus on standards for the end-to-end VR ecosystem. VRIF has released Guidelines [12], which cover the entire VR distribution ecosystem, including compression, storage and delivery, and aim at documenting the best practices for VR content production and distribution as well as advocating interoperability and deployment guidelines. When it comes to compression, storage and distribution of VR content, VRIF Guidelines recommend the use of selected OMAF media profiles and toolset brands. VRIF Guidelines are continually updated, and the latest update, i.e., version 2.3, added new video profiles and toolset brands from OMAF 2nd edition.

VRIF Guidelines recommend the following OMAF video profiles:

- HEVC-based viewport-independent OMAF video profile.
- Unconstrained HEVC-based viewport-independent OMAF video profile. This profile is specifically recommended for 8K video.
- HEVC-based viewport-dependent OMAF video profile.
- Simple tiling OMAF video profile. This profile is specifically recommended for 8K video. The VRIF Guidelines describe the following three tiling schemes that can be used for viewport-dependent streaming when the decoding capacity enables 8K: 1) Mixed-quality tiling for monoscopic 8K ERP; 2) Mixed-quality tiling for stereoscopic 6K CMP; 3) Mixed-resolution tiling for stereoscopic effective 8K CMP, where the content of the viewport originates from 8K CMP.

In addition, VRIF Guidelines specify the HEVC-based FOV enhanced video profile, where the video is coded into a low-quality bitstream with the entire content coverage and several high-quality sub-picture bitstreams covering different regions of the content. Each bitstream is encapsulated into a track conforming to the HEVC-based viewport-dependent OMAF video profile. Players can selectively obtain sub-picture bitstreams that cover the viewport in addition to consistent processing of the low-quality full-coverage bitstream. Players are required to run several HEVC video decoders in parallel but do not need to perform tile binding.

VRIF Guidelines recommend the use the OMAF 3D audio baseline media profile and suggest the support of viewpoint and overlay toolset brands as specified in OMAF.

## VI. Conclusion and Outlook

Omnidirectional media consumed on head-mounted displays provides an immersive media experience that is superior to traditional 2D and 3D media. The Omnidirectional MediA Format (OMAF) international standard enables interoperable omnidirectional media applications. OMAF second edition introduces new technologies including enabling multiple viewpoints, support of improved overlay of graphics or textual data on top of omnidirectional video, efficient signaling of videos structured in multiple sub parts, and new profiles supporting dynamic bitstream generation according to the viewport. This paper reviewed the multiple viewpoint and overlay features of OMAF and discussed potential usage scenarios for them. The paper also presented the interoperability points included in OMAF and reviewed which OMAF interoperability points have been taken into use in the 3rd Generation Partnership Project (3GPP) Technical Specification (TS) 26.118 on VR profiles for streaming applications and the Virtual Reality Industry Forum Guidelines. Work is ongoing to add support in OMAF for video and image profiles based on the Versatile Video Coding standard [15], with technical finalization expected in October 2022.